\documentclass[Physsubmission, Phys]{SciPost}

\hypersetup{
    colorlinks,
    linkcolor={red!50!black},
    citecolor={blue!50!black},
    urlcolor={blue!80!black}
}

\usepackage[bitstream-charter]{mathdesign}
\DeclareSymbolFont{usualmathcal}{OMS}{cmsy}{m}{n}
\DeclareSymbolFontAlphabet{\mathcal}{usualmathcal}

\begin{document}

\begin{center}

{\Large \textbf{PMC$_\infty$: Infinite-Order Scale-Setting method
using the Principle of Maximum Conformality and preserving the
Intrinsic Conformality \\}}

\end{center}

\begin{center}

Leonardo Di Giustino\textsuperscript{1$\star$}, Stanley J.
Brodsky\textsuperscript{2$\dag$}, Xing-Gang
Wu\textsuperscript{3$\ddag$} and Sheng-Quan
Wang\textsuperscript{4$\S$}

\end{center}

\begin{center}

{\bf 1} Department of Science and High Technology, University of
Insubria, via valleggio 11, I-22100, Como, Italy \\

{\bf 2} SLAC National Accelerator Laboratory, Stanford University,
Stanford, California 94039, USA Department of Physics \\

{\bf 3} Chongqing University, Chongqing 401331, P.R. China \\

{\bf 4} Department of Physics, Guizhou Minzu University, Guiyang
550025, P.R. China \\

$\star$email: ldigiustino@uninsubria.it\\ $\dag$email:
sjbth@slac.stanford.edu , $\ddag$email: sqwang@cqu.edu.cn,
$\S$email: wuxg@cqu.edu.cn \\
\end{center}

\begin{center}
\today
\end{center}


\definecolor{palegray}{gray}{0.95}
\begin{center}
\colorbox{palegray}{
  \begin{tabular}{rr}
  \begin{minipage}{0.1\textwidth}
    \includegraphics[width=35mm]{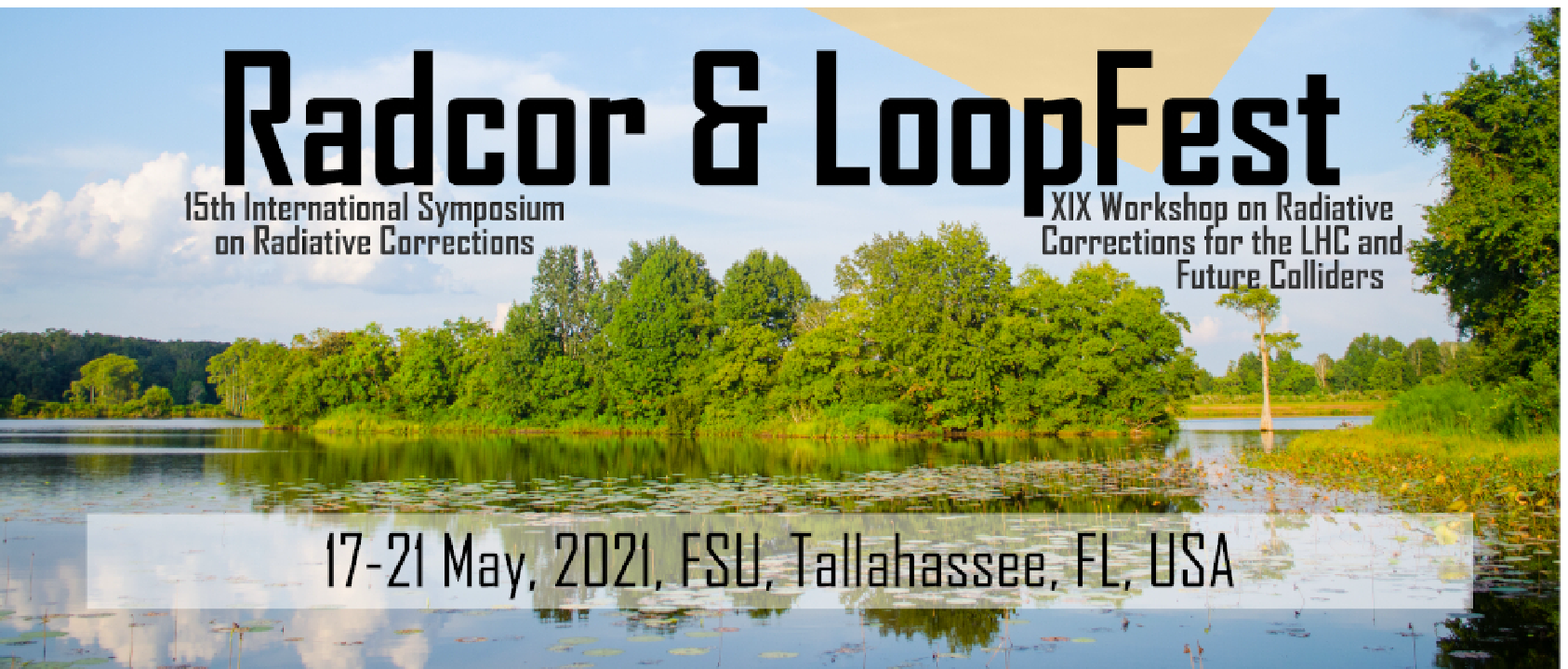}
  \end{minipage}
  &
  \begin{minipage}{0.85\textwidth}
    \begin{center}
    {\it 15th International Symposium on Radiative Corrections: \\Applications of Quantum Field Theory to Phenomenology,}\\
    {\it FSU, Tallahasse, FL, USA, 17-21 May 2021} \\
    \doi{10.21468/SciPostPhysProc.?}\\
    \end{center}
  \end{minipage}
\end{tabular}
}
\end{center}

\section*{Abstract}
{\bf We show results for Thrust and C-parameter in $e^+ e^-$
annihilation to 3 jets obtained using the recently developed new
method for eliminating the scale ambiguity and the scheme
dependence in pQCD namely the Infinite-Order Scale-Setting method
using the Principle of Maximum Conformality (PMC$_\infty$). This
method preserves an important underlying property of gauge
theories:  intrinsic Conformality (iCF).  It leads to a remarkably
efficient method to eliminate the conventional renormalization
scale ambiguity at any order in pQCD.  A comparison with
Conventional Scale Setting method (CSS) is also shown. }

\vspace{10pt}
\noindent\rule{\textwidth}{1pt}
\tableofcontents\thispagestyle{fancy}
\noindent\rule{\textwidth}{1pt}
\vspace{10pt}

\section{Introduction}
\label{sec:intro}

One of the obstacles in making precision tests of the quantum
chromodynamics (QCD) is the uncertainty in setting the
renormalization scale $\mu_R$ into running coupling
$\alpha_s(\mu_R^2)$ for the perturbative expansion of a scale
invariant quantity.

The conventional practice (i.e. conventional scale setting - CSS)
of simply guessing the scale $\mu_R$ of the order of a typical
momentum transfer Q in the process, and then varying the scale
over a range Q/2 and 2Q, leads to predictions that are affected by
large renormalization scale ambiguities.

Additionally, the CSS procedure is not consistent with the
Gell-Mann-Low scheme~\cite{Gell-Mann:1954yli} in Quantum
Electrodynamics (QED)\cite{maitre:2009xp}, the pQCD predictions
are affected by scheme dependence and the resulting perturbative
QCD series is also factorially divergent like
$n!\beta^n_0\alpha^n_s$, i.e. the "renormalon"
problem~\cite{Beneke:1998ui}. Given the factorial growth, the hope
to suppress  scale uncertainties by including higher-order
corrections is compromised.  We recall that there is no ambiguity
in setting the renormalization scale in QED. The standard
Gell-Mann-Low scheme determines the correct renormalization scale
identifying the scale with the virtuality of the exchanged photon.
For example, in electron-muon elastic scattering, the
renormalization scale is the virtuality of the exchanged photon,
i.e. the spacelike momentum transfer squared $\mu_R^2 = q^2 = t$.
Thus
\begin{equation} \alpha(t) = {\alpha(t_0) \over 1 - \Pi(t,t_0)}
\label{qed1}
\end{equation} where
$$ \Pi(t,t_0) = {\Pi(t) -\Pi(t_0)\over 1-\Pi(t_0) } $$
From Eq.\ref{qed1} it follows that the renormalization scale
$\mu_R=t$ can be determined by the $\beta_0$-term and it sums up
all the vacuum polarization contributions into the dressed photon
propagator, both proper and improper at all orders. Given that the
pQCD and pQED predictions match analytically in the $N_C\to 0$
limit where $C_F\alpha_{QCD} \to \alpha_{QED}$ (see
ref.~\cite{Brodsky:1997jk}) it would be convenient to extend the
same procedure to pQCD. A solution to the scale ambiguity problem
is offered by the {\emph{Principle of  Maximum Conformality}}
(PMC)~\cite{Brodsky:1982gc,Brodsky:2011ig,Brodsky:2011ta,
Brodsky:2012rj, Mojaza:2012mf, Brodsky:2013vpa}. This method
provides a systematic way to eliminate renormalization
scheme-and-scale ambiguities from first principles by absorbing
the $\beta$ terms that govern the behavior of the running coupling
via the renormalization group equation. Thus, the divergent
renormalon terms cancel, which improves the convergence of the
perturbative QCD series. Furthermore, the resulting PMC
predictions do not depend on the particular scheme used, thereby
preserving the principles of renormalization group
invariance~\cite{Brodsky:2012ms, Wu:2014iba}. The PMC procedure is
also consistent with the standard Gell-Mann-Low method in the
Abelian limit, $N_c\rightarrow0$~\cite{Brodsky:1997jk}. Besides,
in a theory of unification of all forces, electromagnetic, weak
and strong interactions, such as the Standard Model, or Grand
Unification theories, is highly desirable to use only one method.
The PMC offers the possibility to apply the same method in all
sectors of a theory, starting from first principles, eliminating
the renormalon growth, the scheme dependence, the scale ambiguity,
and satisfying the QED Gell-Mann-Low scheme
in the zero-color limit $N_c\to 0$.\\
The recently developed PMC$_\infty$: {\it Infinite-Order
Scale-Setting using the Principle of Maximum
Conformality}~\cite{DiGiustino:2020fbk} is a new method based on
the PMC and it preserves the property that we define as
\textit{Intrinsic Conformality} (\textit{iCF}). This property
stems directly from an analysis of the perturbative QCD
corrections and leads to scale invariance of an observable
calculated at any fixed order independently from the particular
process or kinematics. Here we apply this method to the Event
Shape Variables : Thrust and C-parameter, showing results and
comparison with the CSS.


\section{The Thrust and C-parameter at NNLO and the CSS}

The thrust ($T$) and C-parameter ($C$) are defined as
\begin{eqnarray}
T=\max\limits_{\vec{n}}\left(\frac{\sum_{i}|\vec{p}_i\cdot\vec{n}|}{\sum_{i}|\vec{p}_i|}\right),
\end{eqnarray}
\begin{eqnarray}
C=\frac{3}{2}\frac{\sum_{i,j}|\vec{p_i}||\vec{p_j}|\sin^2\theta_{ij}}{\left(\sum_i|\vec{p_i}|\right)^2},
\end{eqnarray}
where $\vec{p}_i$ denotes the three-momentum of particle $i$. For
the thrust, the unit vector $\vec{n}$ is varied to define the
thrust axis $\vec{n}_T$ by maximizing the sum on the right-hand
side. For the C-parameter, $\theta_{ij}$ is the angle between
$\vec{p_i}$ and $\vec{p_j}$. It is often used the variable
$(1-T)$, which for the LO of the 3 jet production is restricted to
the range $(0<1-T<1/3)$ and for the C-parameter is $0\leq C\leq
0.75$. (For a review on the Event Shape variables see Refs.
\cite{Ellis:1980wv,Kunszt:1980vt, Vermaseren:1980qz,
Fabricius:1981sx, Giele:1991vf,
Catani:1996jh,Gehrmann-DeRidder:2007nzq,GehrmannDeRidder:2007hr,
Ridder:2014wza, Weinzierl:2008iv, Weinzierl:2009ms,Abbate:2010xh,
Banfi:2014sua}.)

 In general a normalized IR safe single variable observable, such as
the thrust or the C-parameter distribution for the $e^+
e^-\rightarrow 3jets$ \cite{DelDuca:2016ily,DelDuca:2016csb}, is
given by the sum of pQCD contributions calculated up to NNLO at
the initial renormalization scale $\mu_0$:
\begin{eqnarray}
\frac{1}{\sigma_{tot}} \! \frac{O d \sigma(\mu_{0})}{d O}\! & = &
\left\{ x_0 \cdot \frac{ O d \bar{A}_{\mathit{O}}(\mu_0)}{d O} +
x_0^2 \cdot \frac{ O d \bar{B}_{\mathit{O}}(\mu_0)}{d O}  +
x_0^{3} \cdot \frac{O d\bar{C}_{\mathit{O}}(\mu_0)}{d O}+ {\cal
O}(\alpha_{s}^4) \right\},
 \label{observable1}
\end{eqnarray}
where $x(\mu)\equiv \alpha_s(\mu)/(2\pi)$, $O$ is the selected Event Shape variable, $\sigma$ the cross section of the process,\\
$$\sigma_{tot}=\sigma_{0} \left( 1+x_0
A_{t o t}+ x_0^{2} B_{t o t}+ {\cal
O}\left(\alpha_{s}^{3}\right)\right)$$ \\ is the total hadronic
cross section and $\bar{A}_O, \bar{B}_O, \bar{C}_O$ are
respectively the normalized LO, NLO and NNLO coefficients:
\begin{eqnarray}
\bar{A}_{O} &=&A_{O} \nonumber \\
\bar{B}_{O} &=&B_{O}-A_{t o t} A_{O} \\
\bar{C}_{O} &=&C_{O}-A_{t o t} B_{O}-\left(B_{t o t}-A_{t o
t}^{2}\right) A_{O}. \nonumber
\end{eqnarray}
where $A_O, B_O, C_O$ are the coefficients normalized to the tree
level cross section $\sigma_0$ calculated by MonteCarlo (see e.g.
EERAD and Event2 codes~\cite{Gehrmann-DeRidder:2007nzq,
GehrmannDeRidder:2007hr, Ridder:2014wza, Weinzierl:2008iv,
Weinzierl:2009ms}) and $A_{\mathit{tot}}, B_{\mathit{tot}}$ are:
\begin{eqnarray}
A_{\mathit{tot}} &= & \frac{3}{2} C_F ; \nonumber \\
B_{\mathit{tot}} &= & \frac{C_F}{4}N_c +\frac{3}{4}C_F
\frac{\beta_0}{2} (11-8\zeta(3)) -\frac{3}{8} C_F^2.
 \label{norm}
\end{eqnarray}
where $\zeta$ is the Riemann zeta function.

In general according to CSS the renormalization scale is set to
$\mu_0=\sqrt{s}=M_{Z_0}$ and theoretical uncertainties are
evaluated using standard criteria. In this case, we have used the
definition given in Ref.~\cite{Gehrmann-DeRidder:2007nzq} of the
parameter $\delta$, we define the average error for the event
shape variable distributions as:
 \begin{equation}
\bar{\delta}=\frac{1}{N} \sum_i^N \frac{
\text{max}_{\mu}(\sigma_i(\mu))- \text{min}_{\mu}
(\sigma_i(\mu))}{2 \sigma_i(\mu=M_{Z_0})} , \label{delta}
\end{equation}
where $i$ is the index of the bin and $N$ is the total number of
bins, the renormalization scale is varied in the range: $\mu \in
[M_{Z_0}/2; 2 M_{Z_0}]$.

\section{The iCF : conformal coefficients and intrinsic scales}

We define \textit{Intrinsic Conformality} as the unique property
of a renormalizable SU(N)/U(1) gauge theory, like QCD, which
yields to a particular structure of the perturbative corrections
that can be made explicit representing the perturbative
coefficients of Eq. \ref{observable1} using the following RG
invariant parametrization:
\begin{eqnarray}
 A_{O}(\mu_0)\!\!\! &=& \!\!\! A_{\mathit{Conf}} , \nonumber \\
B_{O}(\mu_0) \!\!\! &=& \!\!\! B_{\mathit{Conf}}+\frac{1}{2} \beta_{0} \ln \left(\frac{\mu_0^{2}}{\mu_{I}^{2}}\right) A_{\mathit{Conf}},  \nonumber \\
 C_{O}(\mu_0)\!\!\! &=& \!\!\! C_{\mathit{Conf}} +\beta_{0} \ln \left(\frac{\mu_{0}^{2}}{\mu_{II}^{2}}\right)B_{\mathit{Conf}} + \frac{1}{4}\left[\beta_{1}+\beta_{0}^{2}
 \ln \left(\frac{\mu_0^{2}}{\mu_{I}^{2}}\right)\right] \ln \left(\frac{\mu_0^{2}}{\mu_{I}^{2}}\right)
 A_{\mathit{Conf}}\nonumber \\
 \label{newevolution}
  \end{eqnarray}
where the $A_{\mathit{Conf}}, B_{\mathit{Conf}},
C_{\mathit{Conf}}$ are the scale invariant \textit{Conformal
Coefficients} (i.e. the coefficients of each perturbative order
not depending on the scale $\mu_0$) while we define the $\mu_N$ as
\textit{Intrinsic Conformal Scales} and $\beta_0,\beta_1$ are the
first two coefficients of the $\beta$-function\cite{Gross:1973id,Politzer:1973fx,Caswell:1974gg,Jones:1974mm,Egorian:1978zx}.

By collecting together the terms identified by the same conformal
coefficient, we obtain the observable written in \textit{conformal
subset} ($\sigma_n$) :
\begin{eqnarray}
& \sigma_I &\!\! =\!\!\left\{ \left(\frac{\alpha_{s}(\mu_{0})}{2
\pi}\right) + \frac{1}{2} \beta_{0} \ln \left(
\frac{\mu_0^{2}}{\mu_{I}^{2}}\right)\left(\frac{\alpha_{s}(\mu_0)}{2
\pi}\right)^2  \right.  \nonumber \\ \!\!\! &\!\!\!\!\! +
\!\!\!\!\!& \!\!\!\left. \!\!\!\! \frac{1}{4}\! \left[ \beta_{1}
\! + \! \beta_{0}^{2} \ln
\left(\frac{\mu_0^{2}}{\mu_{I}^{2}}\right) \right] \ln
\left(\frac{\mu_0^{2}}{\mu_{I}^{2}}\right)\!\!
\left(\frac{\alpha_{s}(\mu_{0})}{2 \pi}\right)^3 \!\!\!+\ldots\!\!
\right\}
\!\! A_{\mathit{Conf}}   \nonumber \\
&  \sigma_{II} &\!\! = \!\! \left\{
\left(\frac{\alpha_{s}(\mu_{0})}{2 \pi}\right)^2 \!\!\!\! +\!
\beta_0 \ln \! \left( \! \frac{\mu_0^2}{\mu_{II}^2}\right)\!\!
\left(\frac{\alpha_{s}(\mu_{0})}{2 \pi}\right)^3 \!\!\! +\ldots
\!\! \right\}\!
B_{\mathit{Conf}}   \nonumber \\
& \sigma_{III} &\!\!= \!\! \left\{
\left(\frac{\alpha_{s}(\mu_{0})}{2
\pi}\right)^3 +\ldots \right\}C_{\mathit{Conf}},   \nonumber \\
& \vdots & \qquad
\qquad  .^{.^{.}} \nonumber \\
&  \sigma_{n} &\!\!= \!\!\left\{
\left(\frac{\alpha_{s}(\mu_{0})}{2 \pi}\right)^n
\right\}\mathcal{L}_{n \mathit{Conf}}, \label{confsubsets}
\end{eqnarray}
Any combination of the conformal subsets, $\sigma_{I},
\sigma_{II}, \sigma_{III},...$ such as $\sigma_N=\sum_i \sigma_i$
is still conformal :
\begin{eqnarray}
\left(\mu^2 \frac{ \partial}{\partial \mu^2} +\beta
(\alpha_s)\frac{\partial}{\partial \alpha_s}\right) \sigma_N=0.
\label{sigmainvariance}
\end{eqnarray}

We define here this property of Eq. \ref{confsubsets} of
separating an observable in the union of ordered scale invariant
disjoint subsets $\sigma_{I}, \sigma_{II}, \sigma_{III},...$ as
\textit{ordered scale invariance}.

The coefficients of Eq. \ref{newevolution} can be identified from
a numerical either theoretical perturbative calculation. For the
purpose we use the NNLO results calculated in Refs.
\cite{Weinzierl:2008iv,Weinzierl:2009ms}. Since the leading order
is already ($A_{\mathit{Conf}}$) void of $\beta$-terms we start
with NLO coefficients. A general numerical/theoretical calculation
keeps tracks of all the color factors and the respective
coefficients:
\begin{eqnarray}
B_{O}(N_f)=C_F \left[ C_A B_{O}^{N_c}+C_F B_{O}^{C_F}+ T_F N_f
B_{O}^{N_f}\right] \label{Bcoeff}
\end{eqnarray}
where $C_F=\frac{\left(N_{c}^{2}-1\right)}{2 N_{c}}$, $C_A=N_c$
and $T_F=1/2.$  We can determine the conformal coefficient
$B_{\mathit{Conf}}$ of the NLO order straightforwardly, by fixing
the number of flavors $N_f$ in order to kill the $\beta_0$ term:
\begin{eqnarray}
B_{\mathit{Conf}}&=& B_{O} \left( N_f \equiv \frac{33}{2} \right),\nonumber \\
B_{\beta_0} \equiv  \log  \frac{\mu_0^2}{\mu_I^2}  & = & 2
\frac{B_O-B_{\mathit{Conf}}}{\beta_0 A_{\mathit{Conf}}}
\label{Bconf}
\end{eqnarray}
 we would achieve the same results in the usual PMC
way, i.e. identifying the $N_f$ coefficient with the $\beta_0$
term and then determining the conformal coefficient.  At the NNLO
a general coefficient is made of the contribution of six different
color factors:
\begin{eqnarray}
C_{O}(N_f)&=& \frac{C_F}{4} \left\{ N_{c}^{2} C_{O}^{N_c^2
}+C_{O}^{N_c^0}+\frac{1}{N_{c}^{2}} C_{O}^{\frac{1}{N_c^2}}
 \right. \nonumber \\ & & \!\!\!\!\!\!\!\!\!\!\!\!\!\!\!\!\!\!\!\!\left.\!\!\!\!\!  +N_{f} N_{c}\cdot C_{O}^{N_f N_c}+ \frac{N_{f}}{N_{c}}
C_{O}^{N_f/N_c}+N_{f}^{2} C_{O}^{N_f^2}\right\}. \label{Ccoeff}
\end{eqnarray}
In order to identify all the terms of Eq.\ref{newevolution} we
notice first that the coefficients of the terms $\beta_0^2$ and
$\beta_1$ are already given by the NLO coefficient $B_{\beta_0}$,
then we need to determine only the $\beta_0$- and the conformal
$C_{\mathit{Conf}}$-terms. In order to determine the latter
coefficients we use the same procedure we used for the NLO , i.e.
we set the number of flavors $N_f \equiv 33/2$ in order to drop
off all the $\beta_0$ terms. We have then:
\begin{eqnarray}
C_{\mathit{Conf}}&=& C_{O} \left( N_f \equiv \frac{33}{2} \right)- \frac{1}{4}\overline{\beta}_1 B_{\beta_0} A_{\mathit{Conf}},\nonumber \\
C_{\beta_0} \equiv \log\left(\frac{\mu_0^2}{\mu_{II}^2}\right) & =
& \frac{1}{\beta_0 B_{\mathit{Conf}}} {\bigg (}
C_O-C_{\mathit{Conf}}  \nonumber \\
&\!\!\!\!  &\!\!\!\! -\left.\frac{1}{4} \beta_0^2 B_{\beta_0}^2
A_{\mathit{Conf}}- \frac{1}{4} \beta_1 B_{\beta_0}
A_{\mathit{Conf}} \!\right)\!\!,\nonumber \\
 \label{Cconf}
\end{eqnarray}
with $\overline{\beta}_1\equiv \beta_1(N_f=33/2)=-107$. This
procedure can be extended to all orders and one may also decide
whether to cancel the $\beta_0$, $\beta_1$ or $\beta_2$ by fixing
the appropriate number of flavors.
 We point out that extending the Intrinsic
Conformality to all orders we can predict at this stage the
coefficients of all the color factors of the higher orders related
to the $\beta$-terms except those related to the higher order
conformal coefficients and $\beta_0$-terms (e.g. at NNNLO the
$D_{\mathit{Conf}}$ and $D_{\beta_0}$). The $\beta$-terms are
coefficients that stem from UV-divergent diagrams connected with
the running of the coupling constant and not from UV-finite
diagrams. UV-finite $N_F$ terms may arise but would not contribute
to the $\beta$-terms. These terms should be considered as
conformal terms.

%

\section{The PMC$_\infty$ renormalization scales}

 According to the PMC$_\infty$ , renormalization scales are set to
 the intrinsic scales, and Eq.\ref{observable1} becomes:
\begin{equation}
\frac{1}{\sigma_{tot}} \! \frac{O d
\sigma(\mu_I,\tilde{\mu}_{II},\mu_{0})}{d O}=
\left\{\overline{\sigma}_{I}+\overline{\sigma}_{II}+\overline{\sigma}_{III}+
{\cal O}(\alpha_{s}^4) \right\},
 \label{observable3}
\end{equation}
where the $\overline{\sigma}_{N}$ are normalized conformal subsets
that are given by:
\begin{eqnarray}
\overline{\sigma}_{I} &=& A_{\mathit{Conf}} \cdot x_I \nonumber  \\
\overline{\sigma}_{II} &=& \left( B_{\mathit{Conf}}+\eta
A_{\mathit{tot}} A_{\mathit{Conf}} \right)\cdot x_{II}^2
 - \eta A_{\text{tot}} A_{\mathit{Conf}} \cdot x_0^2 \nonumber \\
 & & -A_{\text{tot}} A_{\mathit{Conf}}\cdot x_0 x_I  \nonumber \\
\overline{\sigma}_{III} &=&\!\! \left( C_{\mathit{Conf}} \!-\!
A_{\text {tot}} \!B_{\mathit{Conf}}\!-\!(B_{\text
{tot}}-A_{\text {tot}}^{2}) A_{\mathit{Conf}}\right) \cdot x_0^3 ,\nonumber \\
\label{normalizedcoeff}
\end{eqnarray}
where $x_I,x_{II},x_0$ are the couplings determined at the
$\mu_I,\tilde{\mu}_{II},\mu_0$ scales respectively.

Normalized conformal subsets for the region $(1-T)>0.33$ and
$C>0.75$ can be achieved simply by setting
$A_{\mathit{Conf}}\equiv 0$ in the Eq. \ref{normalizedcoeff}. The
PMC$_\infty$ scales, $\mu_N$ , are given by:

\begin{eqnarray}
\large{\mu}_{\mathit{I}} & = & \sqrt{s} \cdot e^{f_{sc}-\frac{1}{2} B_{\beta_0}},\hspace{1.9cm}{ \scriptstyle (1-T)<0.33   \; ,\;    C<0.75}  \nonumber \\
\large{\tilde{\mu}}_{\mathit{II}} & =& \left\{ \begin{array}{lr}
\sqrt{s} \cdot e^{f_{sc}-\frac{1}{2} C_{\beta_0} \cdot
\frac{B_{\mathit{Conf}}}{B_{\mathit{Conf}}+\eta \cdot
A_{\mathit{tot}} A_{\mathit{Conf}} }},\\ \hspace{3.9cm} {\scriptstyle (1-T)<0.33 \; , \;  C<0.75} ,   \\
  \sqrt{s}\cdot e^{f_{sc}-\frac{1}{2} C_{\beta_0}},\\ \hspace{3.9cm}{ \scriptstyle (1-T)>0.33  \; , \;    C>0.75}
  \label{icfscale}\\
 \end{array} \right.
 \label{PMC12}
\end{eqnarray}

and $\mu_0=M_{Z_0}$. The renormalization scheme factor for the QCD
results is set to $f_{sc}\equiv 0$.

\begin{figure}[htb]
\centering
\includegraphics[width=0.450\textwidth]{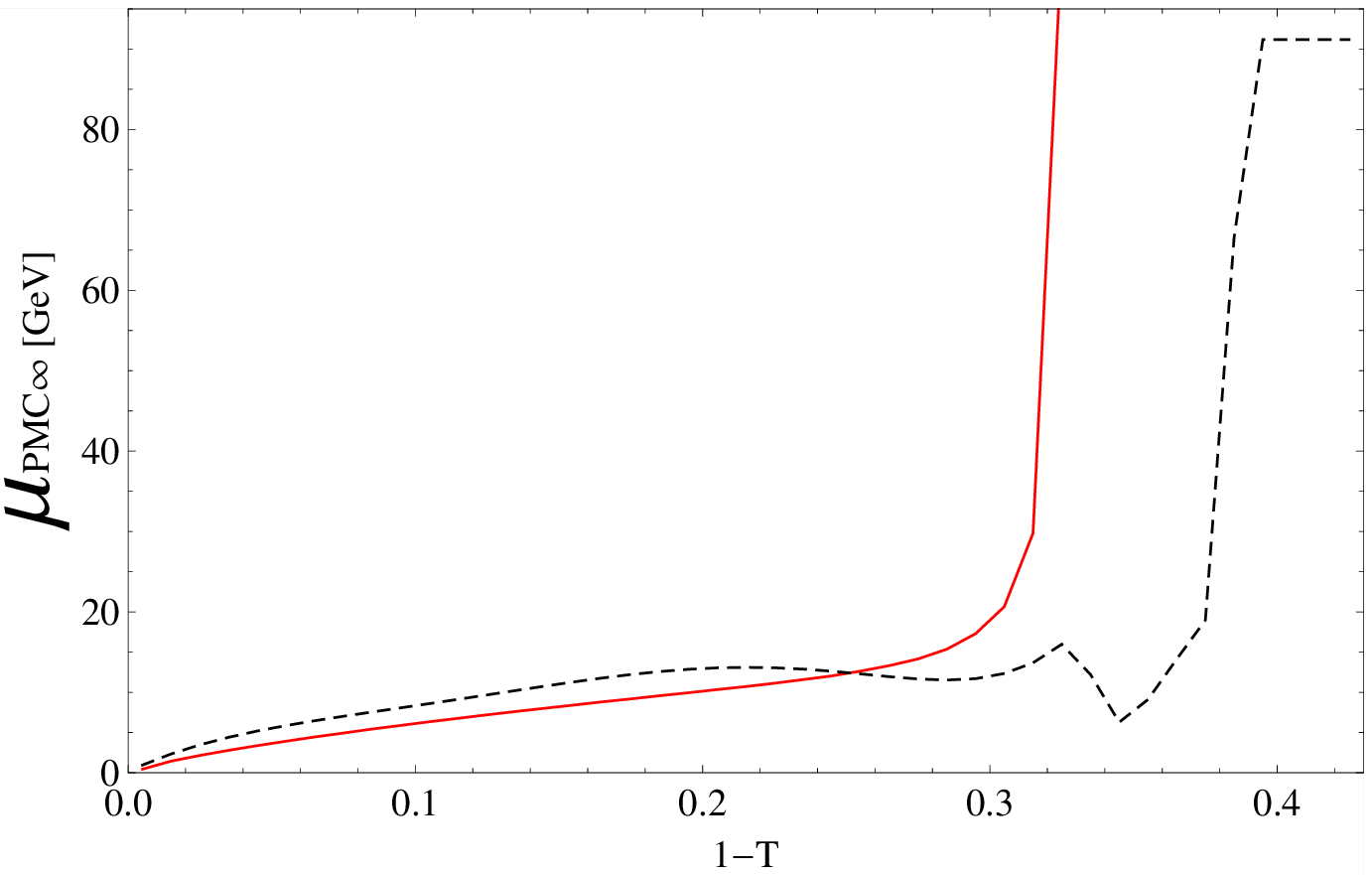}\hspace{0.1cm}
\includegraphics[width=0.460\textwidth]{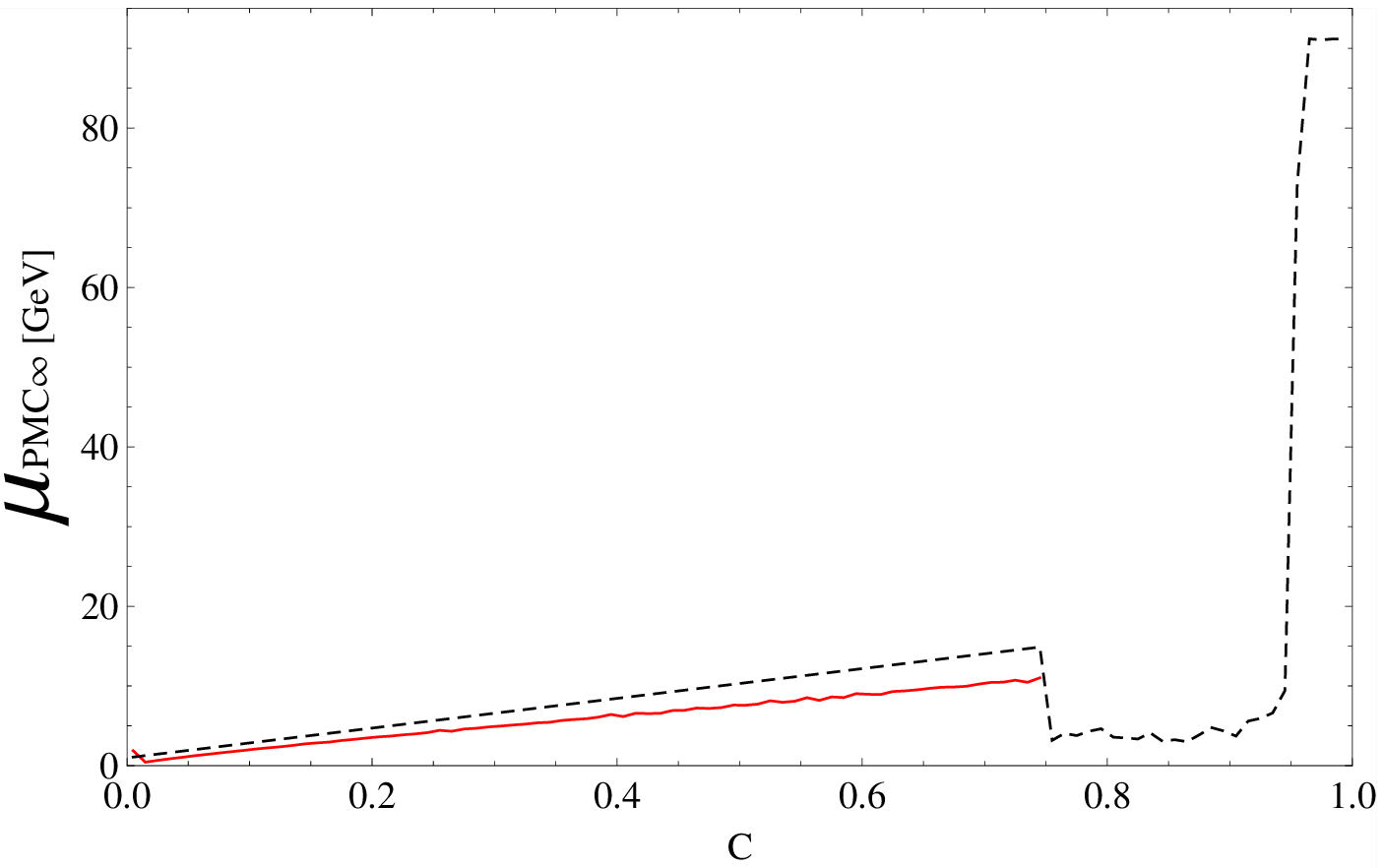}\\
\caption{The LO-PMC$_\infty$ (Solid Red) and the NLO-PMC$_\infty$
(Dashed Black) scales for thrust (Left) and C-parameter (Right). }
\label{Tscales}
\end{figure}

The $\eta$ parameter is a regularization term in order to cancel
the singularities of the NLO scale, $\mu_{II}$, in the range
$(1-T)<0.33$ and $C<0.75$, depending on non-matching zeroes
between numerator and denominator in the $C_{\beta_0}$. In general
this term is not mandatory for applying the PMC$_{\infty}$, it is
necessary only in case one is interested to apply the method all
over the entire range covered by the thrust, or any other
observable. Its value has been determined to $\eta=3.51$ for both
thrust and C-parameter distribution and it introduces no bias
effects up to the accuracy of the calculations and the related
errors are totally negligible up to this stage. The LO and NLO
PMC$_\infty$ scales for thrust and C-parameter are shown in
Fig.\ref{Tscales}.
The PMC$_\infty$ scales are functions of the center-of mass-energy
$\sqrt{s}$  and of the event shape variable. We notice that LO and
NLO PMC$_\infty$ scales have similar behaviors in the range
$(1-T)<0.33$  and $C<0.75$ going to zero at the lower boundary.

\section{Comparison of the CSS and PMC$_\infty$ Results }
\label{sec:comp}

\begin{figure}[htb]
\centering
\includegraphics[width=0.60\textwidth]{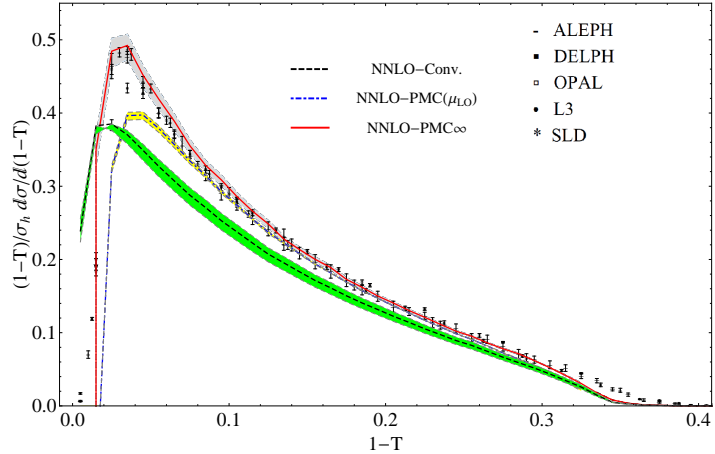}
\caption{The thrust distribution at NNLO under the Conventional
(Dashed Black), the PMC($\mu$\textsubscript{LO}) (DotDashed Blue)
and the PMC$_\infty$ (Solid Red). The experimental data points are
taken from the ALEPH, DELPHI,OPAL, L3, SLD experiments
\cite{aleph,delphi,opal,l3,sld}.  The shaded areas show
theoretical errors predictions at NNLO.} \label{thrust2}
\end{figure}

\begin{figure}[h]
\centering
\includegraphics[width=0.60\textwidth]{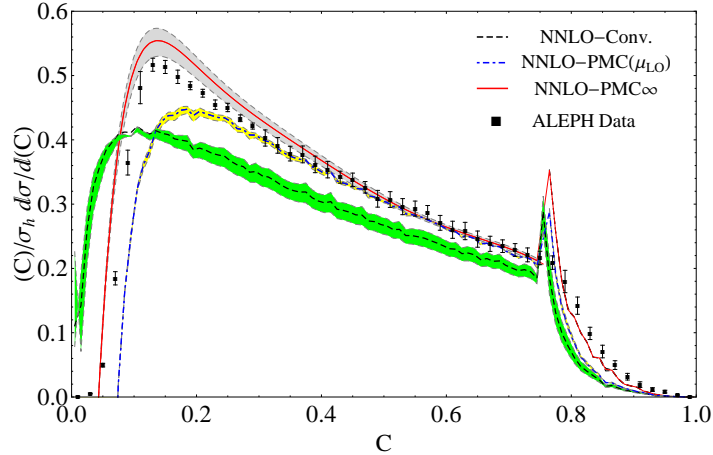}
\caption{The NNLO C-parameter distribution under the Conventional
Scale Setting (Dashed Black), the PMC($\mu$\textsubscript{LO})
(DotDashed Blue ) and the PMC$_\infty$ (Solid Red). The
experimental data points (Black) are taken from the ALEPH
experiment~\cite{aleph}. The shaded area shows theoretical errors
predictions at NNLO.} \label{Cpar2}
\end{figure}

We show in Fig.\ref{thrust2} and Fig. \ref{Cpar2} results for the
thrust and C-parameter with a direct comparison of the
PMC$_\infty$ with the the CSS method. In addition we have shown
also the results of the first PMC approach used in
\cite{Wang:2019ljl,Wang:2019isi} that we indicate as
PMC($\mu$\textsubscript{LO}) extended to the NNLO accuracy. In
this approach the last unknown PMC scale $\mu$\textsubscript{NLO}
of the NLO has been set to the last known PMC scale
$\mu$\textsubscript{LO} of the LO, while the NNLO scale
$\mu$\textsubscript{NNLO}$\equiv \mu_0$ has been set to the
kinematic scale $\mu_0\equiv \sqrt{s}$. This analysis has been
performed in order to show that the procedure of setting the last
unknown scale to the last known one leads to stable and precise
results and is consistent with proper PMC method in a wide range
of values of the $(1-T)$ and $C$ variable. Using the PMC$_\infty$,
average errors in the range $0.<(1-T)<0.42$ of the thrust improve
from $\bar{\delta}\simeq 7.36\%$ to $1.95\%$  and in the range
$0<(C)<1$ of the C-parameter from $\bar{\delta}\simeq 7.26\%$ to
$2.43\%$ from NLO to NNLO respectively. Average errors calculated
in different regions of the spectrum are reported in Table
\ref{tab:1} for thrust and C-parameter. From the comparison with
the CSS we notice that the PMC$_\infty$ prescription significantly
improves the theoretical predictions. Besides, results are in
remarkable agreement with the experimental data in a wider range
of values for both the $1-T$ and $C$ variables and they show an
improvement of the PMC$(\mu_{LO})$ results when the two-jets and
the multi-jets regions are approached, i.e. the region near the
lower and the upper boundary respectively. The use of the
PMC$_\infty$ approach on perturbative thrust QCD-calculations
restores the correct behavior of the thrust distribution in the
region $(1-T)>0.33$ and $C>0.75$ and this is a clear effect of the
iCF property. Comparison with the experimental data has been
improved all over the spectrum and the introduction of the $N^3LO$
order correction would improve this comparison especially in the
multi-jet region. In the PMC$_\infty$ method theoretical errors
are given by the unknown intrinsic conformal scale of the last
order of accuracy. We expect this scale not to be significantly
different from that of the previous orders. In this particular
case, as shown in Eq.\ref{normalizedcoeff}, we have also a
dependence on the initial scale $\alpha_s(\mu_0)$ left due to the
normalization and to the regularization terms. These errors
represent the 12.5\% and 1.5\% respectively of the whole
theoretical errors in the range $0<(1-T)<0.42$ and they could be
improved by means of a correct normalization.

\begin{table}[h!]
\centering
 \begin{tabular}{||c|c|c|c||}
   \hline
 $\bar{\delta}[\%]$  &  CSS  & PMC($\mu$\textsubscript{LO}) & PMC$_\infty$ \\
    \hline

    $0.00 < (1-T) < 0.33$ &  5.34  & 1.33 & 1.77\\
   $0.00 < (1-T) <0.42$ & 6.00 & -  &  1.95 \\
     $0.00 < (C) < 1.00$ & 6.47 & 1.55 & 2.43 \\
   \hline
   \end{tabular}
 \caption{Average error, $\bar{\delta}$, for NNLO Thrust and C-parameter distributions under CSS, PMC($\mu$\textsubscript{LO}) and PMC$_\infty$
 scale settings calculated in different intervals.}
 \label{tab:1}
\end{table}

\section{Conclusion}

In this article we have shown results for thrust and C-parameter
for $e^+e^-\rightarrow 3 jets$, comparing the two methods for
setting the renormalization scale in pQCD: the Conventional Scale
Setting (CSS) and the infinite-order scale setting based on the
Principle of Maximum Conformality (PMC$_\infty$). The PMC$_\infty$
method preserves the unique property of the Intrinsic Conformality
(iCF). This property leads to a RG invariant parametrization which
underlies the {\it ordered scale invariance}. The PMC$_\infty$
method solves the renormalization scale ambiguity, eliminates the
scheme dependence and is consistent with the Gell-Mann and Low
scheme in QED. We point out that in fixed order calculations the
PMC$_\infty$ last scale is set to the kinematic scale of the
process: in this case $\mu_{III}=\sqrt{s}=M_{Z_0}$. As shown in
Eq. \ref{confsubsets}, the scale dependence on the initial scale
is totally confined in the last subset $\sigma_n$. Thus the the
last term in the iCF determines the level of {\it conformality}
reached by the expansion and is entangled with theoretical
uncertainties given by higher order uncalculated terms. Any
variation of the last scale has to be intended to evaluate
theoretical uncertainties given by higher order contributions and
not an ambiguity of the PMC$_\infty$ method
\cite{Chawdhry:2019uuv}. Evaluation of the theoretical errors
using standard criteria shows that the PMC$_\infty$ significantly
improves the precision of the pQCD calculations for thrust and
C-parameter. We remark that an improved analysis of theoretical
errors might be obtained by giving a prediction on the
contributions of higher order terms using a statistical approach
as shown in Ref. \cite{Bonvini:2020xeo,Duhr:2021mfd}. This would
lead to a more rigorous method to evaluate errors and thus to
restrict the range of the last PMC$_\infty$ scale that, as we have
shown here, can also be fixed to the last known PMC$_\infty$ one
leading to precise and stable predictions.

\section*{Acknowledgements}
LDG thanks the organizers of RADCOR 2021 for the opportunity to
make this presentation. This research was supported in part by the
Department of Energy contract {\text DE-AC02-76SF00515} (SJB).
SLAC-PUB-17627

\bibliography{SciPost_Example_BiBTeX_File.bib}

\nolinenumbers

\end{document}